\documentclass[a4paper,10pt,twoside]{cpc-hepnp}

\usepackage{multicol}
\usepackage{graphicx}
\usepackage{booktabs}
\usepackage{amssymb,bm,mathrsfs,bbm,amscd}
\usepackage[tbtags]{amsmath}
\usepackage{lastpage}
\usepackage{tabularx,multirow,array}

\begin{document}
\fancyhead[c]{\small Chinese Physics C~~~Vol. xx, No. x (201x) xxxxxx}
\fancyfoot[C]{\small 010201-\thepage}

\footnotetext[0]{}
\title{Spectroscopy of light $N^{*}$ baryons}

\author{%
      Zalak Shah$^{1)}$\email{zalak.physics@gmail.com}, Keval Gandhi and
\quad Ajay Kumar Rai$^{2)}$\email{raiajayk@gmail.com}
}
\maketitle
\address{
Department of Applied Physics, Sardar Vallabhbhai National Institute of Technology, Surat, Gujarat, India-395007\\}
\begin{abstract}
We present the masses of N baryon upto 3300 MeV. The radial and orbital excited states are determined using hypercentral constituent quark model with first order correction. The obtained masses are compared with experimental results and other theoretical prediction. The Regge Trajectories are also determined in (n, $M^2$) and (J, $M^2$) planes. Moreover, the magnetic moments for $J^{P}= \frac{1}{2}^{+}, \frac{1}{2}^{-}$ are calculated. We also calculate the $N\pi$ decay width of excited nucleons.
\end{abstract}

\footnotetext[0]{\hspace*{-3mm}\raisebox{0.3ex}{$\scriptstyle\copyright$}2013
Chinese Physical Society and the Institute of High Energy Physics
of the Chinese Academy of Sciences and the Institute
of Modern Physics of the Chinese Academy of Sciences and IOP Publishing Ltd}%

\begin{multicols}{2}
\section{Introduction}
The evaluation of the static and dynamic properties of the hadrons have always been a main concern of theoretical and experimental community of nuclear and particle physics. The new experimental observations in the field of light baryons, heavy baryons and exotic states are determined very recently. As the time progresses, the different experimental groups are providing light baryon resonances with increasing confidence level of these states \cite{BelleYelton2018,Aaij2018bottom,Aaij2017ch,olive}. These experimental observations have created renewed interest to study light baryon spectra(radial as well as orbital excited states) theoretically. The  description of the all baryon systems can be found in the review articles\cite{crede,Samios,Klempt2009,Valcarce2005review,gianni,Sonnenschein}.\\

 The combination of three confined light quarks with flavor up, down and strange leads to the study of light baryons, belongs to the $SU(3)_f$ symmetry multiplets and provide a vent to understand the non-abelian character of Quantum Chromodynamics(QCD). The light baryons can be in the following multiplets:	
\begin{center}
$10_S \oplus 8_M \oplus 8_M \oplus 1_A$
\end{center}

In this paper, we study the N$^*$ baryon system which is the combination of two $d$ quarks and one $u$ quark. The nature and interactions of the compound system can be evaluated performing hadron spectroscopy. N$^*$’s can provide us with critical insights into the nature of QCD in the confinement domain and its relevance to nuclear and low energy hadron physics \cite{Isgur:2000ad}. Particle Data Group(PDG) has provided the many excited states resonances of N baryons \cite{olive}. The search for these resonances are the main focuses of the baryon programs at JLab\cite{JLab2012},Mainzer Mikrotron (MAMI)\cite{MAMI2016}, the Beijing Spectrometer (BES), the Electron Stretcher and Accelerator (ELSA) facility (the Crystal Barrel collaboration)\cite{CBELSA2005}, GRAAL \cite{GRAAL2003,GRAAL2017} and the Two Arms Photon Spectrometer(TAPS) \cite{TAPS2000}, SAPHIR\cite{SAPHIR,SAPHIR2004}, CLAS \cite{Mokeev2016CLAS}. We can also expect the new results from the analysis project like EBAC, Julich, SAID, MAID, etc. Many theoretical studies are the phenomenological tool to investigate the excited resonances of nucleons; various quark models   \cite{Eichmann2017,DeSanctis2016,SantopintoPRC2012,Giannini2005,
Giannini2001,Isgur,Capstick,Glozman}, Basis Light Front Quantization (BLFQ) approach \cite{James2018}, BnGa partial wave analysis\cite{Anisovich2012}, chiral Lagrangian theory \cite{Khemchandani2013}, Fadeev approach \cite{Sanchis-Alepuz2014,Valcarce2005} Lattice QCD \cite{Edwards2011,WalkerLoud2008,Lin2008},etc. \\

We determined the spectra of $N^*$ resonances in the framework of hypercentral Constituent Quark Model(hCQM). The model has already been extended to provide the excited state baryon resonances in heavy sector.  The excited states masses of singly, doubly and triply heavy baryons in both charm and bottom sector are in good agreement with other theoretical predictions as well as with the recent experimental outcomes\cite{Shahcpc2018,ZShah2016singly,shah2018few,ZShah2017triply,zcpc,ZShah2017doubly,zbottom}. Further, the magnetic moments of heavy baryons are also determined at L=0 and radiative decays, strong decays are calculated \cite{keval}. Moreover, with the help of the Regge trajectories the unknown quantum states and $J^P$ values are also identified. In this paper, we will use the same model and identify the several hadronic properties of N$^*$ baryons.  Giannini et al. has used this  model to study the light flavored baryons for low lying radial as well as orbital excited states below 2GeV. Where we have calculated the spectra for 1S-5S, 1P-5P, 1D-4D, and 1F-2F states and constructed Regge Trajectories upto 3.3 GeV. It is also very clear from Fig (1) that the new experimental resonances are very near to our predicted states. The brief description of model is given in Section 2. After obtaining masses we also determined the several hadronic properties using these masses. Such as, Regge trajectories and magnetic moments and  some decay widths of excited states are discussed in Section 3 followed by conclusion in Section 4.

\section{The Model}
Spectroscopy has long been a powerful tool for the constituents of a compound system in order to expose their nature and interactions.   In this context, hadron spectroscopy has already produced many surprises. We have determined the heavy baryon spectroscopy which means the singly, doubly and triply heavy baryons excited states starting from S state to F states using hypercentral Constituent Quark Model(hCQM) with all isospin splittings; in our previous work \cite{Shahcpc2018,ZShah2016singly,shah2018few,ZShah2017triply,zcpc,ZShah2017doubly,zbottom}.
\\

\begin{center}
\tabcaption{\label{tab:table1} The list of the N resonances from Baryon Summary Table of PDG(2018).}
\begin{tabular*}{80mm}{c@{\extracolsep{\fill}}cccccccc}
\toprule
State&$J^{P}$&M(Exp.)&status\\
\noalign{\smallskip}\hline\noalign{\smallskip}
N(939)&$\frac{1}{2}^{+}$&939&****\\
N(1440)&$\frac{1}{2}^{+}$&1420-1470&****\\
N(1710)&$\frac{1}{2}^{+}$&1680-1740&***\\
N(2100)&$\frac{1}{2}^{+}$&2100&*\\
N(1535)&$\frac{1}{2}^{-}$&1525-1545&****\\
N(1520)&$\frac{3}{2}^{-}$&1515-1525&****\\
N(1895)&$\frac{1}{2}^{-}$&1880-1910&**\\
N(1680)&$\frac{5}{2}^{+}$&1680-1690&****\\
N(1875)&$\frac{3}{2}^{-}$&1820-1920&***\\
N(2000)&$\frac{5}{2}^{+}$&1950-2150&**\\
N(1720)&$\frac{3}{2}^{+}$&1700-1750&****\\
N(1880)&$\frac{1}{2}^{+}$&-&**\\
N(1990)&$\frac{7}{2}^{+}$&1950-2100&**\\
N(2190)&$\frac{7}{2}^{-}$&2100-2200&****\\
N(2250)&$\frac{9}{2}^{-}$&2250-2320&****\\
N(2570)&$\frac{5}{2}^{-}$&-&**\\
\hline
\end{tabular*}
\end{center}

In the present study, we predict radial and orbital excited states of $N$ baryon in the hypercentral approach\cite{Santopinto1996,gianni}.  
The two relative coordinates are rewritten into a single six dimensional vector and the non-relativistic {\it Schr$\ddot o$dinger} equation. We solve six dimensional space equation numerically in Mathematica notebook\cite{lucha}. The potential expressed in terms of the hypercentral radial co-ordinate, takes care of the three body interactions effectively. We employ the  Coulomb plus linear confinement potential with first order correction for the quarks. The relative Jacobi coordinates can be expressed as \cite{Bijker2000}

\begin{subequations}
\begin{equation}
\vec{\rho} = \dfrac{1}{\sqrt{2}}(\vec{r_{1}} - \vec{r_{2}}),
\end{equation}
\begin{equation}
\vec{\lambda} =\dfrac{m_1\vec{r_1}+m_2\vec{r_2}-(m_1+m_2)\vec{r_3}}{\sqrt{m_1^2+m_2^2+(m_1+m_2)^2}}
%\vec{\lambda}= \dfrac{1}{\sqrt{6}}(\vec{r_1}+ \vec{r_2}- 2\vec{r_3}).
\end{equation}
\end{subequations}
The  Hamiltonian of the baryonic system in the hCQM is then expressed as
\begin{equation}
H=\dfrac{P_{x}^{2}}{2m} +V(x).
\end{equation}where, $m=\frac{2 m_{\rho} m_{\lambda}}{m_{\rho} + m_{\lambda}}$, is the reduced mass and $x$ is the six dimensional radial hyper central coordinate of the three body system.  The quark masses are $m_u$=$m_d$=0.290 GeV. We add first  order correction to the energy term of potential to observed the effect. In heavy sector baryons, we demonstrate results without adding correction as well as with first order correction. But, here, we only consider the excited state masses adding first order correction. The model is formed by a linear confining interaction with a spin, flavor and orbital angular momentum dependent hyperfine interaction. The potential is in form of,

\begin{equation}\label{eq:7}
V(x) =  V^{0}(x) + \left(\dfrac{1}{m_{\rho}}+ \dfrac{1}{m_{\lambda}}\right) V^{(1)}(x)+V_{SD}(x).
\end{equation}
where $V^{0}(x)$ and first order correction $V^{1}(x)$ is given by
\begin{equation}
V^{(0)}(x)= \dfrac{\tau}{x}+ \beta x,
\quad
V^{(1)}(x)= - C_{F}C_{A} \dfrac{\alpha_{s}^{2}}{4 x^{2}}.
\end{equation}
$V^{(0)}(x)$ is the sum of hyper Coulomb(hC) (vector) interaction and a confinement (scalar) term. Here, the hyper-Coulomb strength $\tau = -\frac{2}{3} \alpha_{s}$ where $\alpha_{s}$ corresponds to the strong running coupling constant;  $\beta$ corresponds to the string tension of the confinement. $C_{F}$ and $C_{A}$ are the Casimir charges of the fundamental and adjoint representation. \\

For computing the mass difference between different degenerate baryonic states, we consider the spin dependent part of the usual one gluon exchange potential (OGEP). The spin-dependent part, $V_{SD}(x)$ contains three types of the interaction terms, such as the spin-spin term $V_{SS} (x)$, the spin-orbit term $V_{\gamma S}(x)$ and tensor term $V_{T}(x)$ given by,
\begin{eqnarray}
\label{eq:VSD}
V_{SD}(x)= V_{SS}(x)(\vec{S_{\rho}}.\vec{S_\lambda})
+ V_{\gamma S}(x) (\vec{\gamma} \cdot \vec{S}) + V_{T} (x)\\ \nonumber
\left[ S^2-\dfrac{3(\vec{S }\cdot \vec{x})(\vec{S} \cdot \vec{x})}{x^{2}} \right]
\end{eqnarray}
The coefficient of these spin dependent terms of can be written as
\begin{subequations}
\begin{equation}
V_{\gamma S} (x) = \dfrac{1}{2 m_{\rho} m_{\lambda}x}  \left(3\dfrac{dV_{V}}{dx} -\dfrac{dV_{S}}{dx} \right),
\end{equation}
\begin{equation}
V_{T}(x)=\dfrac{1}{6 m_{\rho} m_{\lambda}} \left(3\dfrac{d^{2}V_{V}}{dx^{2}} -\dfrac{1}{x}\dfrac{dV_{V}}{dx} \right),
\end{equation}
\begin{equation}
V_{SS}(x)= \dfrac{1}{3 m_{\rho} m_{\lambda}} \bigtriangledown^{2} V_{V}.
\end{equation}
\end{subequations}
the Spin-orbit and the tensor term describe the fine structure of the states, while the spin- spin term gives the spin singlet triplet splittings.  

\end{multicols}

\begin{table*}

%\begin{center}
\tabcaption{\label{tab:table2} The predicted excited state masses of N baryon(in MeV).}
\begin{tabular}{cccccccccccccc}
\toprule
State&$J^{P}$&Mass&Exp. \cite{olive}&\cite{zahraepjp}&\cite{Giannini2005}&\cite{Santopinto2015}&\cite{Giannini2005}&\cite{salehi2012}&\cite{Chen2008}&\cite{Aslanzadeh2017}&\cite{Anisovich2012}&\cite{Isgur}&\cite{Capstick}\\
\hline
1S&$\frac{1}{2}^{+}$&939&938&938&938&939&938&938&939&938&960\\
2S&$\frac{1}{2}^{+}$&1425&1420-1470&1444&1448&1511&1463&1467&1440&1492&1430\\
3S&$\frac{1}{2}^{+}$&1721&1680-1740&1832&1795&1776&1752&1710&1710&1763&1710\\
4S&$\frac{1}{2}^{+}$&2089&2100&&&&&&2100\\
5S&$\frac{1}{2}^{+}$&2515\\
\hline
1P&$\frac{1}{2}^{-}$&1565&1525-1545&1567&1543&1537&1524&&1535&1511&1501&1490&1460\\
1P&$\frac{3}{2}^{-}$&1535&1515-1525&1567&1543&1537&1524&1536&1520&1511&1517&1535&1495\\
1P&$\frac{5}{2}^{-}$&1495&&&&&&&&&&&1630\\
\noalign{\smallskip}\hline
2P&$\frac{1}{2}^{-}$&1898&1880-1910&&&1888&&&1650&&1895\\
2P&$\frac{3}{2}^{-}$&1865&1820-1920&&&&&&1700&&1880\\
2P&$\frac{5}{2}^{-}$&1820&&&&&&&1675\\
\noalign{\smallskip}\hline
3P&$\frac{1}{2}^{-}$&2288&&&&&&&2090\\
3P&$\frac{3}{2}^{-}$&2251& 2150&&&&&&2080&&2150\\
3P&$\frac{5}{2}^{-}$&2202\\
\noalign{\smallskip}\hline
4P&$\frac{1}{2}^{-}$&2741\\
4P&$\frac{3}{2}^{-}$&2697\\
4P&$\frac{5}{2}^{-}$&2628\\
\hline
5P&$\frac{1}{2}^{-}$&3242\\
5P&$\frac{3}{2}^{-}$&3192\\
5P&$\frac{5}{2}^{-}$&3126\\\hline
1D&$\frac{1}{2}^{+}$&1849&1835-1910&&&1890&&&&&1870\\
1D&$\frac{3}{2}^{+}$&1815&1700-1750&&&1648&&&&1735&1690\\
1D&$\frac{5}{2}^{+}$&1769&1680-1690&&	&1799&1680&1704&&1735&1689\\
1D&$\frac{7}{2}^{+}$&1712\\
\hline
2D&$\frac{1}{2}^{+}$&2244\\
2D&$\frac{3}{2}^{+}$&2204\\
2D&$\frac{5}{2}^{+}$&2150&1950-2150&&&&&&&&2090\\
2D&$\frac{7}{2}^{+}$&2083&1950-2100&&&&&&&&2060\\
\hline
3D&$\frac{1}{2}^{+}$&2694\\
3D&$\frac{3}{2}^{+}$&2648\\
3D&$\frac{5}{2}^{+}$&2586\\
3D&$\frac{7}{2}^{+}$&2510\\
\hline
4D&$\frac{1}{2}^{+}$&3197\\
4D&$\frac{3}{2}^{+}$&3144\\
4D&$\frac{5}{2}^{+}$&3074\\
4D&$\frac{7}{2}^{+}$&2986\\
\hline
1F&$\frac{3}{2}^{-}$&2167\\
1F&$\frac{5}{2}^{-}$&2112\\
1F&$\frac{7}{2}^{-}$&2045&2100-2200&&&&&&2190&&2180\\
1F&$\frac{9}{2}^{-}$&1963\\
\hline
2F&$\frac{3}{2}^{-}$&2614\\
2F&$\frac{5}{2}^{-}$&2551\\
2F&$\frac{7}{2}^{-}$&2473\\
2F&$\frac{9}{2}^{-}$&2379&2250-2320&&&&&&&&2280\\
\hline
\end{tabular}
%\end{center}

\end{table*}

\begin{figure*}
\includegraphics[scale=0.7]{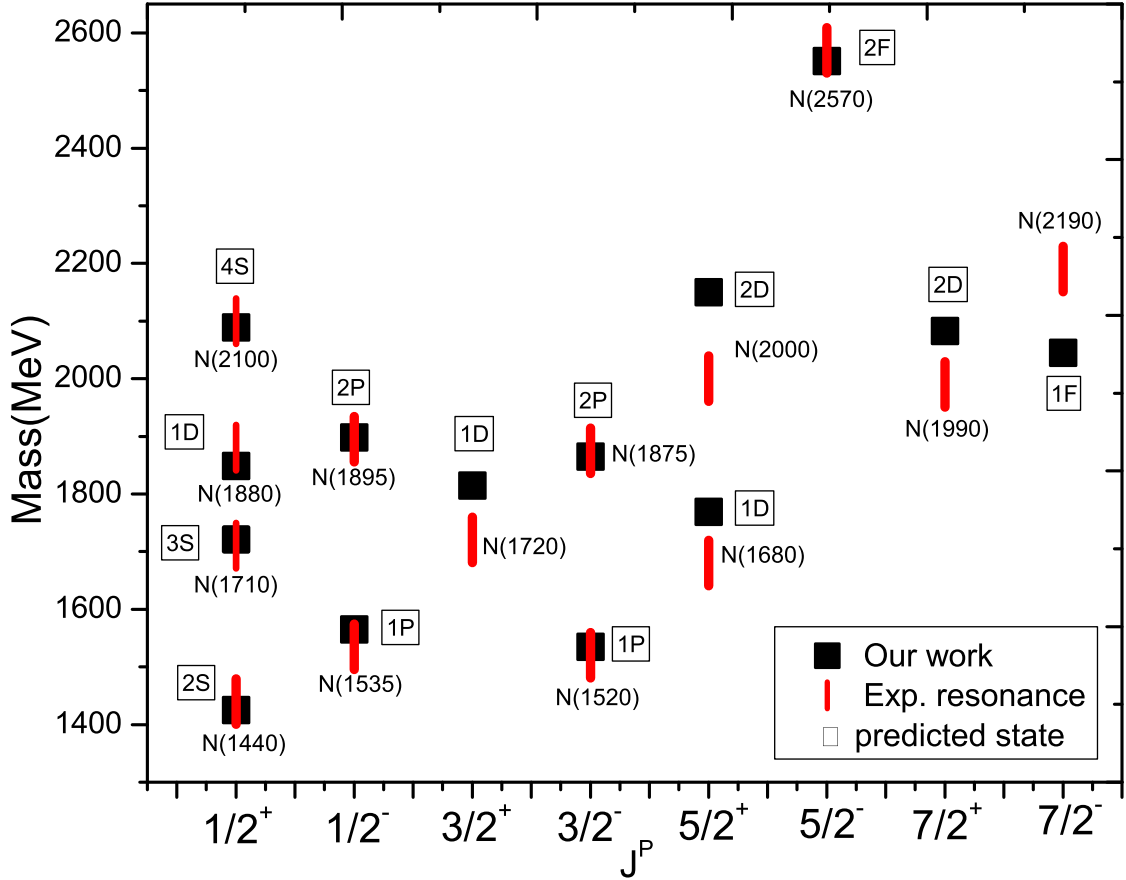}% Here is how to import EPS art
\caption{\label{fig:1}Light baryon classification is presented. Square represents our predicted masses with particular state and bar shows experimental available resonances. }
\end{figure*}

\begin{multicols}{2}

\section{Results and Discussions}
\subsection{$N^*$ resonances}
N baryon has isospin I= $\frac{1}{2}$ and strangeness $S$= 0. We have constructed a complete classification of $N^{*}$ baryons spectrum in this article. Indeed, many experimentally known excited state resonances fit well with our predicted states. Theses resonances are listed in Table \ref{tab:table1} with their mass, experimental status and respective $J^{P}$ values. The status(*) conveys that 4 stars-existence is certain, 3 star-likely to certain, though more information desirable, 2 star-evidence is fair, 1 star existence is poor. Besides, some more three and four star baryons are also mentioned in PDG(2018) \cite{olive}, such as, N(1650), N(1675), N(1700), N(1900) with  $J^{P}$ values $\frac{1}{2}^{-}$, $\frac{5}{2}^{-}$, $\frac{3}{2}^{-}$, $\frac{3}{2}^{+}$ and the two star baryons N(1860), N(2120), N(2300); all of them do not fit with our obtained resonances.\\

Table \ref{tab:table2} represents the light baryon masses from 1S-5S, 1P-5P, 1D-4D and 1F-2F states. The obtained $N$ baryon masses are tabulated with experimental resonances as well as other theoretical approaches. The radial excited states are calculated for  $J^{P}= \frac{1}{2}^{+}$ and the orbital excited P, D and F states are calculated for $J^{P}$ values, ($\frac{1}{2}^{-}$, $\frac{3}{2}^{-}$, $\frac{5}{2}^{-}$), ($\frac{1}{2}^{+}$, $\frac{3}{2}^{+}$, $\frac{5}{2}^{+}$, $\frac{7}{2}^{+}$) and ($\frac{3}{2}^{-}$, $\frac{5}{2}^{-}$, $\frac{7}{2}^{-}$, $\frac{9}{2}^{-}$), respectively.

\begin{table*}
\centering
\tabcaption{\label{tab:table3}Magnetic moment(in nuclear magneton) for $N(939)$ baryon.}
\begin{tabular}{cccccc}
\toprule
wave-function& Our &Exp.& Ref. \cite{THAKKAR2011}&Ref. \cite{Wang2008mag}&Ref. \cite{Kerbikov2000}\\
\hline
$ \frac{4}{3} \mu_d - \frac{1}{3} \mu_u$
&-1.997&-1.913&-2.07&-1.97&-1.69\\
\hline
\end{tabular}
\end{table*}

The states, $N(1440)$, $N(1710)$, $N(1535)$, $N(1520)$ are listed as a 4-star states. They are determined as 2S, 3S, 1P($\frac{1}{2}^{-}$), 1P($\frac{3}{2}^{-}$) states, also by many other theoretical approaches. The all masses are near to the experimental ranges. We suggest baryons resonances $N(1895)$, $N(1875)$ should be 2P states with $J^{P}$ vales $\frac{1}{2}^{-}$ and $\frac{3}{2}^{-}$. Ref. \cite{Santopinto2015} mass show only 10 MeV difference with our mass of 2P.  Further, the calculated masses of 3P state show $\approx$200 MeV difference with Ref. \cite{Chen2008}.

The assigned $J^{P}$ value of $N(1720)$ is $\frac{3}{2}^{+}$. Our 1D ($\frac{3}{2}^{+}$) state mass is 65 MeV higher than the upper bound of the experimental mass and 80 MeV higher than Ref. \cite{Aslanzadeh2017}, so that we predict as 1D state. Our predicted 1D ($\frac{5}{2}^{+}$) state mass is $\approx$ 80 MeV higher than baryon $N(1680)$ mass and also show the difference of 30, 65, 34 MeV with Refs. \cite{Santopinto2015,salehi2012,Aslanzadeh2017}

The other two baryon resonances $N(1990)$, $N(2000)$ have $J^{P}$ values $\frac{5}{2}^{+}$ and $\frac{7}{2}^{+}$, respectively and our 2D state masses with same $J^{P}$ assignment fits well with them.

Moving towards the higher excited negative parity resonances which are $N(2190)$, $N(2250)$ and $N(2570)$ baryons. Our 1F state mass is 55 MeV less than experimental mass while  2F ($\frac{9}{2}^{-}$) state is 59 MeV higher than experimental mass. The mass of baryon $N(2570)$ is experimentally unknown, if we consider it as 2570 MeV than it shows only 20 MeV difference with our 2F state with $J^{P}=\frac{5}{2}^{-}$.

We do not find any results which gives the masses for higher negative and positive parity excited states (which can be determines as 4P, 5P, 2D, 3D, 4D, 2F states). The $N$ resonances are compared with experimental masses and possible $J^P$ values in Fig. \ref{fig:1}

\subsection{Regge Trajectories}

\begin{figure*}
\centering
\begin{minipage}[b]{0.50\linewidth}
\includegraphics[scale=0.5]{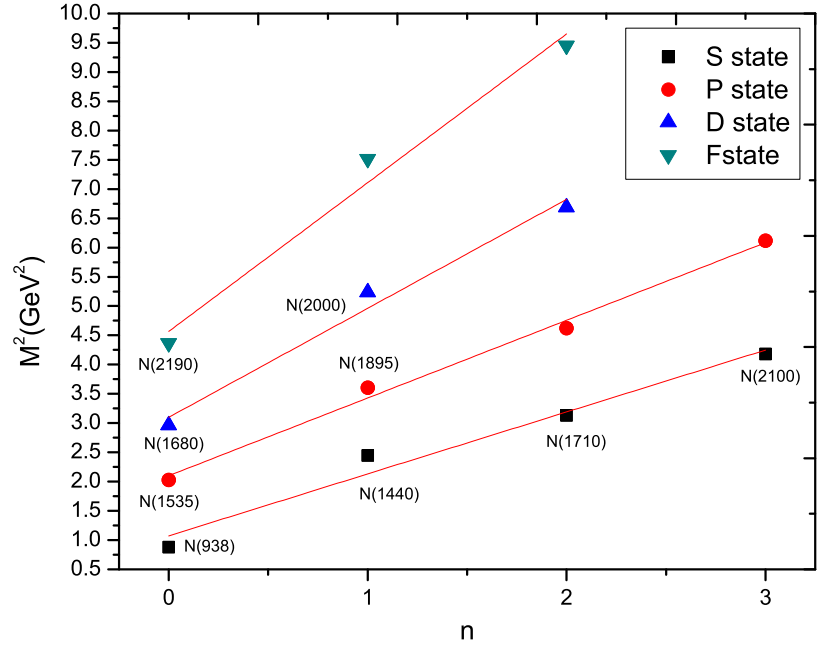}
\label{fig:2}
%\end{minipage}
%\caption{\label{fig:epsart} Regge trajectories (n, $M^2$) plane }
%\end{figure*} 
%
%\begin{figure*}
%\centering
%\begin{minipage}[b]{0.50\linewidth}
\includegraphics[scale=0.5]{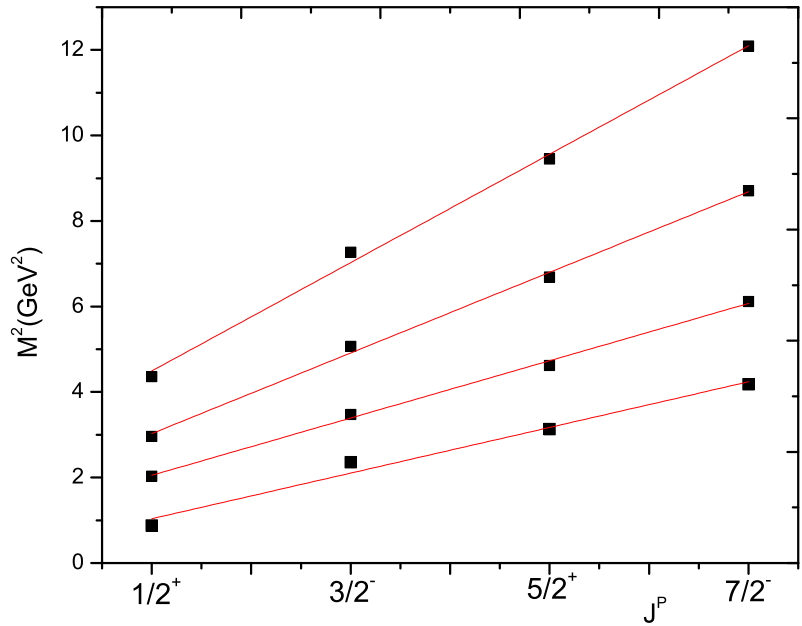}
\label{fig:3}
\end{minipage}
\caption{\label{fig:epsart1} Regge trajectories of N$^*$ baryons in (n, $M^2$) and (J, $M^2$) planes }
\end{figure*}

One  more important property of hadronic spectroscopy are Regge trajectories. Perception of the Regge trajectories is useful in spectral as well as many non-spectral purposes. The spin and mass of the hadrons are related in these plots.  Using the obtained results of Table \ref{tab:table2}, we plot the graphs in (n, $M^{2}$) and (J, $M^{2}$) planes [See Fig. $\ref{fig:epsart1}$].  The relation is,

\begin{equation}
n= \beta M^2+ \beta_0~~~~\&~~~~~~
J= \alpha M^2+ \alpha_0
\end{equation}
where n is a principal quantum number, $\beta$, $\alpha$ are slopes and $\beta_0$, $\alpha_0$ are intercepts. The ground  and radial excited states S (with $J^{P}=\frac{1}{2}^{+}$) and the orbital excited state P (with $J^{P}= \frac{1}{2}^{-}$), D (with $J^{P}= \frac{5}{2}^{+}$) and F (with $J^{P}= \frac{7}{2}^{-}$) are plotted  from bottom to top in ($M^2\rightarrow$ n) plot. In ($M^2\rightarrow$ J) plot, we use ($J^{P}=\frac{1}{2}^{+}$, $J^{P}=\frac{3}{2}^{-}$, $J^{P}=\frac{5}{2}^{+}$, $J^{P}=\frac{7}{2}^{-}$) for the value of principal quantum number, n=1 to 4.  We have already introduced these plots and determined the experimentally unknown heavy baryons in our previous work \cite{Shahcpc2018,ZShah2016singly,
shah2018few,ZShah2017triply,zcpc,
ZShah2017doubly,zbottom}. Now, we want to observe the nature of the graphs in the case of light baryons as well.
%\begin{figure*}
%\centering
%\begin{minipage}[b]{0.50\linewidth}
%\includegraphics[scale=0.50]{ccb_n.eps}
%\label{fig:1}
%\end{minipage}
%\caption{\label{fig:epsart} Regge trajectory in (n, $M^2$) plane.}
%\end{figure*}

%\begin{figure}
%\includegraphics[scale=0.33]{fig_2}
%\caption{\label{fig:2}Regge trajectories in (n, $M^{2}$) plane. }
%\end{figure}
%
%
%\begin{figure}
%\includegraphics[scale=0.33]{fig_3}
%\caption{\label{fig:3}Regge trajectories in (J, $M^{2}$) planes. }
%\end{figure}

\subsection{Magnetic moments}
To emphasis the magnetic moment of nucleon is also an important hadronic properties, as it is very important input for electrmagnetic transitions, form factor and radiative decays of baryons. We here determined the magnetic moment of $N(939)$ state with $J^{P}=\frac{1}{2}^{+}$ and $N(1535)$ states with $J^{P}=\frac{1}{2}^{-}$. The magnetic moments of baryons are obtained in terms of the spin, charge and effective mass of the bound quarks as in Ref.\cite{Shahcpc2018}
\begin{eqnarray}
\mu_{B}=\sum_{i}\langle \phi_{sf}\vert \mu_{iz}\vert\phi_{sf}\rangle);~~~
\mu_{i}=\frac{e_i \sigma_i}{2m_{i}^{eff}},
\end{eqnarray}
where $e_i$ is a charge and $\sigma_i$ is the spin of the respective constituent quark.  The effective mass for each of the constituting quarks $m_{i}^{eff}$ can be defined as
\begin{equation}
m_{i}^{eff}= m_i\left( 1+ \frac{\langle H \rangle}{\sum_{i} m_i} \right)
\end{equation}
where $\langle H \rangle$ = E + $\langle V_{spin} \rangle$. The wave-function and obtainedmagneticmoment is given in Table \ref{tab:table3}.\\

The magnetic moment is also carried out by orbital excitation. The final spin flavor wave function from quark model for the nucleons with $J^{P}= \frac{1}{2}^{-}$ are given by \cite{Nsharma2013} as:

\begin{eqnarray}\nonumber
\left( \frac{1}{9}  \mu_{u}+ \frac{2}{9} \mu_{d} \right)cos^{2} \theta  +\left( \frac{1}{3}  \mu_{u}+  \mu_{d}\right) sin^{2} \theta  \\ \nonumber
-\left(-\frac{8}{9}  \mu_{u}+\frac{8}{9} \mu_{d} \right)cos \theta sin \theta
\end{eqnarray}

\begin{eqnarray}\nonumber
\left( \frac{2}{9}  \mu_{u}+ \frac{1}{9} \mu_{d} \right)cos^{2} \theta  +\left( \frac{8}{9}  \mu_{u}+ \frac{1}{3} \mu_{d}\right) sin^{2} \theta  \\ \nonumber
-\left(\frac{8}{9}  \mu_{u}-\frac{8}{9} \mu_{d} \right)cos \theta sin \theta
\end{eqnarray}
 The value of mixing angle $\theta$= -31.7. The obtained magnetic moment value for N(1535) is -0.7695 and 1.8299, respectively(in nuclear magneton) and they are -1.284 and 1.894 in ref.\cite{Nsharma2013} . The other Refs. \cite{Liu2005,Chiang2002} obtained the values for $J^P= \frac{1}{2}^{-}$ is -0.9 and -0.56 , respectively.

\subsection{Decay width}

The  wave functions of nucleons will be bilinear combinations of spin– flavor and orbital wave functions and then the product of two Jacobi
coordinates of the 3-quark system.  S. Capstick et al. had already determined the decay widths of light baryons in their work \cite{capstickdecay}. The decay width relations will depend on the orbital wave functions of baryons. The decay widths are expressed as in Refs. \cite{Riska2001}.  We adopt the pion-nucleon coupling constants($f$)  from \cite{Riska2001} and determined the widths using our masses as an input.  The Energy($E$) and the momentum of pion $\kappa$ are,
\begin{subequations}
\begin{equation}
E=\frac{m^{* 2}- m_\pi^2+ m_N^2}{2m^*},
\end{equation}

\begin{equation}
\kappa=\frac{\sqrt{[m^{* 2}-(m_N+ m_\pi)^{2}][m^{* 2}-(m_N- m_\pi)^{2}]}}{2m^*}
\end{equation}
\end{subequations}

where, $m_N$= 939MeV and $m_\pi$= 139MeV  are the mass of nucleon and pion respectively. $m^*$ is a mass of resonance used from Table \ref{tab:table2} for each case. We calculate some of the decay widths of the excited nucleons, they are,
\begin{enumerate}
\item $N(1440) \rightarrow N\pi$
\begin{equation}
\Gamma = \frac{3f^2}{4\pi} \frac{E-m_N}{m^*} \frac{\kappa}{m_\pi^{2}}(m^*+m_N)^{2}
\end{equation}
is 62\% which is in the range of PDG 55-75\%  with $f$=0.39 and $m^*$=1425.
\item $N(1535) \rightarrow N\pi$
\begin{equation}
\Gamma = \frac{f^2}{4\pi} \frac{E+m_N}{m^*} \frac{\kappa}{m_\pi^{2}}(m^*-m_N)^{2}
\end{equation}
is 86\% which is higher than the range of PDG 55-65\% and 25-65\% of Ref.\cite{chun2009} with $f$=0.36 and $m^*$=1565.
\item $N(1520) \rightarrow N\pi$
\begin{equation}
\Gamma =\frac{1}{3} \frac{f^2}{4\pi} \frac{E-m_N}{m^*} \frac{\kappa^3}{m_\pi^{2}}
\end{equation}
is 16\% which is lower than range of PDG 32-52\%  with $f$=1.56 and $m^*$=1535.
\item $N(1720) \rightarrow N\pi$
\begin{equation}
\Gamma = \frac{1}{3}\frac{f^2}{4\pi} \frac{E+m_N}{m^*} \frac{\kappa^{3}}{m_\pi^{2}}
\end{equation}
is 11\% which is in the range of PDG 8-14\%  with $f$=0.25 and $m^*$=1815.
\item $N(1680) \rightarrow N\pi$
\begin{equation}
\Gamma = \frac{2}{5}\frac{f^2}{4\pi} \frac{E-m_N}{m^*} \frac{\kappa^{5}}{m_\pi^{4}}
\end{equation}
is 118\% which is higher than  the range of PDG 60-70\%  with $f$=0.42 and $m^*$=1769.
\end{enumerate}

\section{Conclusions}

$N^{*}$ resonances are determined using the hCQM model by adding the  first order correction to the potential. The complete mass spectra with individual states and predicted $J^{P}$ values are presented in Table \ref{tab:table2} and compared with experimental states graphically in Fig. \ref{fig:1}. We can observe that the resonances are predicted from first radial excited state (2S) to the orbital excited state (2F). The important points are concluded as below:
\begin{itemize}
\item  {Among all, we compare 14 experimentally known states with our prediction. We also plot their masses against their $J^{P}$ values. The conclusion of the plot would suggest that the states $N(1440)$, $N(1710)$, $N(1880)$, $N(2100)$, $N(1535)$, $N(1895)$, $N(1520)$, $N(1875)$, $N(2570)$ are very near to our determined resonances. While the states, $N(1720)$, $N(1680)$, $N(2000)$, $N(1990)$, $N(2190)$ are far from our determination. The all determined states are presented according to our predictions of isospin splitting states.}
\item {The radial excited states(2S-4S) masses are within the range of an experimental evidence.}
\item{It can be concluded that the mass spectra of hadrons can be described conveniently through Regge trajectories. These trajectories will guide to identify the quantum number of particular resonance state. We can see in Fig. $\ref{fig:epsart1}$, the trajectories are linear but not parallel same as in Ref. \cite{Masjuan2017}. }

\item{The magnetic moment is calculated for $J^{P}=\frac{1}{2}^{+}$ and $J^{P}=\frac{1}{2}^{-}$.  We will also try to reproduce the nucleon magnetic moments of other excited states after this  preliminary calculation. }

\item{Nucleons $N(1440)$, $N(1535)$, $N(1520)$, $N(1720)$, $N(1680)$ for $N\pi$ decay widths are calculated. In case of $N(1680)$, the ratio of $\frac{\Gamma_{N\pi}}{\Gamma_{tot}} $ is very high. In other cases, it is near/in range.}

\item{We have successfully study the mass spectra of N$^*$ baryons in hypercentral quark model (hCQM)and we would like to extend for other light baryons as well.}

\end{itemize}
%
%dfhwjnfcek
%\bibliographystyle{epj}

%
\end{multicols}
%%
%%
%%\vspace{-1mm}
%%\centerline{\rule{80mm}{0.1pt}}
%%\vspace{2mm}
%%
%%
%%
%%
%%
%%
%%
\begin{multicols}{2}
%%%

%%%\bibitem{valcarce} A. Valcarce, H. Garcilazo, and J. Vijande, Eur. Phys. J. A \textbf{37}, 217 (2008)
%%%\bibitem{9} R. Bijker, F. Iachello and E. Santopinto, J Phys. A \textbf{31}, 9041 (1998);R. Bijker, F. Iachello, and A. Leviatan, Ann. Phys. (N.Y.) 236, 69 (1994).
%%%\bibitem{6} L. Ya. Glozman and D. O. Riska, Phys. Rep. C \textbf{268}, 263 (1996).
%%%
%%%\bibitem{28}  R. Roncaglia, D. B. Lichtenberg, and E.Predazzi, Phys. Rev. D \textbf{52}, 1722 (1995) .
%%%\bibitem{5} M. Ferraris, M. M. Giannini, M. Pizzo, E. Santopinto and L. Tiator, Phys. Lett. B \textbf{364}, 231 (1995).
%%%\bibitem{100} E. E. Jenkins, Phys. Rev. D \textbf{54}, 4515 (1996).
%%%\bibitem{99} L.-H. Liu, L.-Y. Xiao, and X.-H. Zhong, Phys. Rev. D \textbf{86}, 034024 (2012).
%%%\bibitem{98} H.-X. Chen, W. Chen, Q. Mao, A. Hosaka, et al. Phys. Rev. D \textbf{91}, 054034 (2015).
%%%\bibitem{90} J. P. Blanckenberg, and H. Weigel Phys. Lett. B \textbf{750}, 230 (2015).

\end{multicols}
\end{document}